\documentclass[10pt]{llncs}
\usepackage{psfig}

\newcommand{\avida}{{\sf avida}}
\newcommand{\Avida}{{\sf Avida}}
\newcommand{\tierra}{{\sf tierra}}

\begin{document}
\title{Evolution of Differentiated Expression Patterns in Digital Organisms}
\author{Charles Ofria$^\star$, Christoph Adami$^\dagger$, Travis C. Collier$^\ddagger$, and Grace K. Hsu} 
\institute{Beckman Institute\\ California Institute 
           of Technology, Pasadena, California 91125, USA\\
           $^\dagger$W.K. Kellogg Radiation Lab 106-38,
            Caltech, Pasadena, CA 91125\\
           $^\ddagger$Division of Organismic Biology,
            Ecology, and Evolution,\\ UCLA, Los Angeles, CA 90095\\
           \mbox{}\\
           $^\star$ To whom correspondence should be addressed.
           E-Mail: charles@gg.caltech.edu\\
}
\date{\today}

\maketitle

\begin{abstract}
  We investigate the evolutionary processes behind the development and
  optimization of multiple threads of execution in digital organisms
  using the {\avida} platform, a software package that implements
  Darwinian evolution on populations of self-replicating computer
  programs.  The system is seeded with a linearly executed ancestor
  capable only of reproducing its own genome, whereas its underlying
  language has the capacity for multiple threads of execution (i.e.,
  simultaneous expression of sections of the genome.)  We
  witness the {\it evolution} to multi-threaded organisms and track
  the development of distinct expression patterns.  Additionally, we
  examine both the evolvability of multi-threaded organisms and the
  level of thread differentiation as a function of environmental
  complexity, and find that differentiation is more pronounced in
  complex environments.
\end{abstract}

\section{Introduction}

Evolution has traditionally been a formidable subject to study due to
its gradual pace in the natural world.  One successful method uses
microscopic organisms with generational times as short as an hour, but
even this approach has difficulties; it is still impossible to perform
measurements without disturbing the system, and the time-scales to see
significant adaptation remain on the order of weeks, at
best\footnote{Populations of {\it E.coli} introduced into new
  environments begin adaptation immediately, with significant results
  apparent in a few weeks~\cite{LEN3}.}.  Recently, a new tool has
become available to study these problems in a computational
medium---the use of populations of self-replicating computer programs.
These ``digital organisms'' are limited in speed only by the computers
used, with generations in a typical trial taking a few seconds.

Of course, many differences remain between digital and simple
biochemical life, and we address one of the critical ones in this
paper.  In nature, many chemical reactions and genome expressions
occur simultaneously, with a system of gene regulation guiding their
interactions.  However, in digital organisms only one instruction is
executed at a time, implying that no two sections of the program can
directly interact.  Due to this, an obvious extension is to examine
the dynamics of adaptation in artificial systems that have the
capacity for more than one thread of execution (i.e., an independent
CPU with its own instruction pointer, operating on the same genome).

Work in this direction began in 1994 with Thearling and Ray using the
program {\tierra}~\cite{RAY91}.  These experiments were initialized
with an ancestor that creates two threads each copying half of its
genome, thereby doubling its replication rate.  Evolution then
produces more threads up to the maximum allowed~\cite{RAY94}.  In
subsequent papers~\cite{RAY97,RAY98} this research extended to
organisms whose threads are not performing identical operations.  This
is done in an enhanced version of the {\tierra} system ({\bf ``Network
  Tierra}''~\cite{RAY95}), in which multiple ``islands'' of digital
organisms are processed on real-world machines across the Internet.
In these later experiments, the organisms exist in a more complex
environment in which they have the option of seeking other islands on
which to place their offspring.  The ancestor used for these
experiments reproduces while searching for better islands using
independent threads.  Thread differentiation persists only when
island-jumping is actively beneficial; that is, when a meaningful
element of complexity is present in the environment.

In experiments reported on here, we survey the initial emergence of
multiple threads and study their subsequent divergence in function.
We then investigate the hypothesis that environmental complexity plays
a key role in the pressure for the thread execution patterns to
differentiate.

\section{Experimental Details}

We use the {\avida} platform to examine the development of
multi-threading in populations exposed to different environments at
distinct levels of complexity, comparing them to each other and to
controls that lack the capacity for multiple threads.

\subsection{The {\Avida} Platform}

{\Avida} is an auto-adaptive genetic system designed for use as a
platform in Artificial Life research.  The {\avida} system comprises a
population of self-reproducing strings of instructions that adapt to
both an intrinsic fitness landscape (self-reproduction) and an
externally imposed (extrinsic) bonus structure provided by the
researcher.

A standard {\avida} organism is a single genome composed of a sequence
of instructions that are processed as commands to the CPU of a virtual
computer.  This genome is loaded into the memory space of the CPU, and
the execution of each instruction modifies the state of that CPU.  In
addition to the memory, a virtual CPU has three integer registers, two
integer stacks, an input/output buffer, and an instruction pointer.
In standard {\avida} experiments, an organism's genome has one of 28
possible instructions at each line.  The virtual CPUs are
Turing-complete, and therefore do not explicitly limit the ability for
the population to adapt to its computational world.  For more details
on {\avida}, see~\cite{OBA98}.

To allow different sections of a program to be executed in parallel,
we have implemented three new instructions.  A new thread of execution
is initiated with {\tt fork-th}.  This thread has its own registers,
instruction pointer, and a single stack, all initialized to be
identical to the spawning thread.  The second stack is shared to
facilitate communication among threads.  Only the new thread will
execute the instruction immediately following the {\tt fork-th}; the
original will skip it enabling the threads to act and adapt
independently.  If, for example, a jump instruction is at this
location, it may cause the new thread to execute a different section
of the program ({\it segregated differentiation}), whereas a
mathematical operation could modify the outcome of subsequent
calculations ({\it overlapping differentiation}).  On the other hand,
a no-operation instruction at this position allows the threads to
progress identically ({\it non-differentiated}).  We have also
implemented {\tt kill-th}, an instruction that halts the thread
executing it, and {\tt id-th}, which places a unique thread
identification number in a register, allowing the organism to
conditionally regulate the execution of its genome.
 
We performed experiments on three environments of differing
complexity, with both the extended instruction set that allows
multiple expression patterns and the standard instruction set as a
control.  As individual trials can differ extensively in the course of
their evolution, each setup was repeated in two hundred trials to gain
statistical significance.  The experiments were performed on
populations of 3600 digital organisms for 50,000 updates\footnote{An
  update represents the execution of an average of 30 instructions per
  program in the population.  50,000 updates equates to approximately
  9000 generations and takes about 20 hours of execution on a Pentium
  Pro 200. The data and complete genomes are available at {\tt
    http://www.krl.caltech.edu/avida/pubs/ecal99/}\ . }.  Mutations are
set at a probability of 0.75\% for each instruction copied, and a 5\%
probability for an instruction to be inserted or removed in the genome
of a new offspring.

The first environment (I) is the least complex, with no explicit
environmental factors to affect the evolution of the organisms; that
is, the optimization of replication rate is the only adaptive pressure
on the population.  The next environment (II), has collections of
numbers that the organisms may retrieve and manipulate.  We can view
the successful computation of any of twelve logical operations that we
reward\footnote{The completion of a logic operation involves the
  organism drawing one or more 32-bit integers from the environment,
  computing a bitwise logical function using one or more {\tt nand}
  instructions, and outputting the result back into the environment.}
as beneficial metabolic chemical reactions, and speed-up the virtual
CPU accordingly; more complex tasks result in larger speed-ups.  If
the speed increase is more than the time expended to perform the task,
the new functionality is selected for.  The final environment (III)
studied is the most complex, with 80 logic operations rewarded.

A record is maintained of the development of the population, including
the genomes of the most abundant organisms.  For each trial, these
dominant genomes are analyzed to produce a time series of thread use
and differentiation.

\subsection{Differentiation Metrics}

The following measures and indicators keep track of the functional
differentiation of codes.  We keep this initial analysis manageable by
setting a maximum of two threads available to run simultaneously.  The
relaxation of this constraint does lead to the development of more
than two threads with characteristically similar interactions.

{\bf Thread Distance} measures the spatial divergence of the two
instruction pointers.  This measurement is the average {\it distance}
(in units of instructions) between the execution positions of the
individual threads.  If this value becomes high relative to the length
of the genome, it is an indication that the threads are segregated,
executing different portions of the genome at any one time, whereas if
it is low, they likely move in lock-step (or sightly offset) with
nearly identical executions.  Note, however, that if two instruction
pointers execute the code offset by a fixed number of instructions,
but otherwise identically, the thread distance is an inflated measure
of differentiation because the temporal offset does not translate into
differing functionality.

{\bf Code Differentiation} distinguishes execution patterns with
differing {\em behavior}.  A count is kept of how often each thread
executes each portion of the genome.  The code differentiation is the
fraction of instructions in the genome for which these counts differ
between threads.  Thus, this metric is insensistive to the ordering of
execution.

{\bf Execution Differentiation} is a more rigorous measure than code
differentiation.  It uses the same counters, taking into consideration
the {\it difference} in the number of times the threads execute each
instruction.  Thus, if one thread executes a line 5 times and the
other executes it 4 times, it would not contribute as much towards
differentiation as an instruction executed all 9 times by one thread,
and not at all by the other.  This metric totals these differences in
execution counts at each line and then divides the sum by the total
number of multi-threaded executions.  Thus, if the threads are
perfectly synchronized, there is zero execution differentiation, and
if only one thread exclusively executes each line, this metric is
maximized at one.  An execution differentiation of 0.5 indicates that
half of the instructions did not have matched executions in each
thread.

\section{Evolution of Multi-Threaded Organisms}

For our initial investigations, we focus on the 200 trials in
environment III (the most complex), with the extended instruction set,
allowing for multi-threading.

\subsection{Emergence of Multiple Execution Patterns}

Describing a universal course of evolution in any medium is 
not feasible due to the numerous random and contingent factors that
play key roles.  However, there are a number of distinct trends, which
will be discussed further.

\begin{figure}[bt]
  \centerline{\psfig{figure=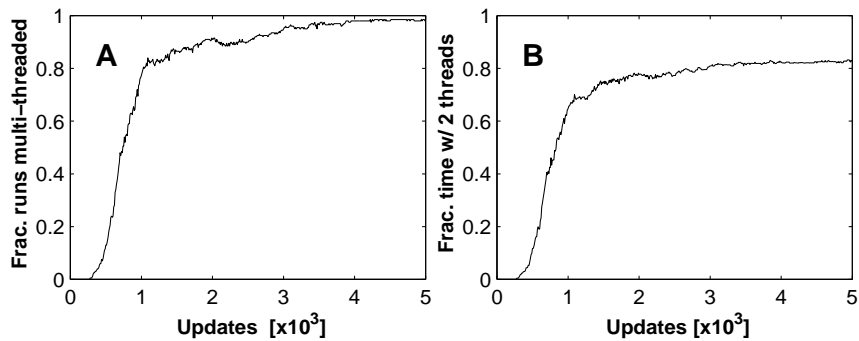,width=4.5in,angle=0}}
 \caption{The time progression of organisms learning to use
   multiple threads averaged over 200 trials.  (A) The fraction of trials
   which thread at all, and (B) The average fraction of time organisms
   spend using both threads at once.  The data displayed here is for the
   first 5000 updates of 50,000 update experiments in environment III.}
 \label{firstfig}
\end{figure}

Let us first consider the transition of organisms from a purely linear
execution to the use of multiple threads.  In Fig.~\ref{firstfig}A, we
see that most populations do develop a secondary thread near the
beginning of their evolution.  Secondary threads come into use as soon
as they grant any benefit to the organisms.  The most common way this
occurs is by having a {\tt fork-th} and a {\tt kill-th} appear around
a section of code, which the threads thereby move through in
lock-step, performing computations twice.  Multiple completions of a
task provide only a minor speed bonus, but this is often sufficient to
warrant a double execution.

Once multiple execution has set in, it will be optimized with time.
Smaller blocks of duplicated code will be expanded, and larger
sections will be used more productively, sometimes even shrinking to
improve efficiency.  Once multiple threads are in use, differentiation
follows.
 
\subsection{Execution Patterns in Multi-threaded Organisms}

\begin{figure}[tb]
 \centerline{\psfig{figure=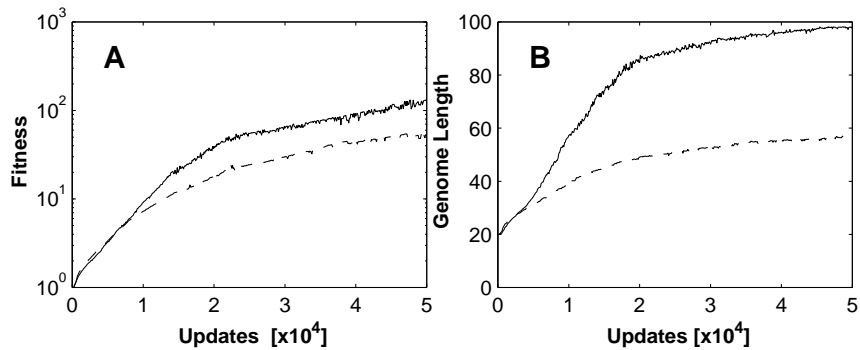,width=4.5in,angle=0}}
 \caption{(A) Average fitness as a function of time (in updates) for the 200
  environment III trials. Most increases to fitness occur as a multiplicative
  factor, requiring fitness to be displayed on a logarithmic scale.
  (B) Average sequence length for the linear
  execution experiments (Solid line) and the multiple execution experiments
  (dashed line).}
 \label{statsfig}
\end{figure}

A critical question is ``What effect does a secondary thread have on
the process of evolution?''  The primary measure to denote a genome's
level of adaptation to an environment is its {\it fitness}.  The
fitness of a digital organism is measured as the number of offspring
it produces per unit time, normalized to the replication rate of the
ancestor.  In all experiments, the fitness of the dominant genotype
starts at one and increases as the organisms adapt.  Fitness
improvements come in two forms: the maximization of CPU speed by task
completion, and the minimization of gestation time.  As all tasks must
be computed each gestation cycle to maintain a reward, this gestation
time minimization includes the {\it optimization} of tasks in addition
to speed-ups in the replication process.  The average progression of
fitness with time is shown in Fig.~\ref{statsfig}A for both the niche
with the expanded instruction set that allows multiple threads, and
the standard, linear execution niche as a control.

Contrary to expectations, the niche that has additional threads
available gives rise to a slower rate of adaptation.  However, the
average length of the genomes (Fig.~\ref{statsfig}B) reveals that the
code for these marginally less fit organisms is stored using 40\%
fewer instructions, indicating a denser encoding.  Indeed, the very
fact that multi-threading develops spontaneously implies that it is
beneficial.  How then can a beneficial development be detrimental to
an organism's fitness?

\begin{figure}[tb]
  \centerline{\psfig{figure=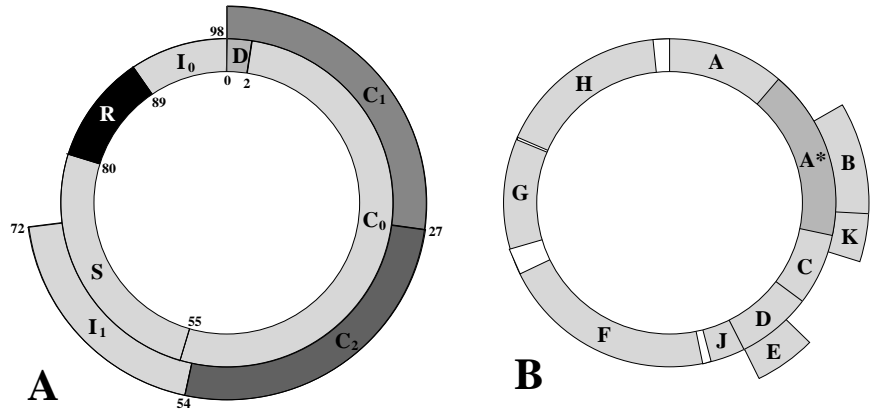,width=4.5in,angle=0}}
 \caption{ {\bf A:} Execution patterns for an evolved {\avida} genome.  The
   inner ring displays instructions executed by the initial thread,
   and the outer ring by the secondary thread.  Darker colors indicate
   more frequent execution. {\bf B:} Genome structure of the phage
   $\Phi$X174.  The promoter sequence for gene $A*$ is entirely within
   gene $A$, causing the genes to express the same series of amino
   acids from the portion overlapped. Genes $B$, $E$, and $K$ are also
   entirely contained within others, but with an offset reading frame,
   such that different amino acids are produced.}
 \label{mapfig}
\end{figure}

Inspection of evolved genomes has allowed us to determine that this
code compression is accomplished by overlapping execution patterns
that differ in their final product.  Fig.~\ref{mapfig}A displays an
example genome.  The initial thread of execution (the inner ring)
begins in the $D$ ``gene'' and proceeds clockwise.  The execution of
$D$ divides the organism when it has a fully developed copy of itself
ready.  This is not the case for this first execution, so the gene
fails with no effect to the organism.  Execution progresses into gene
$C_0$ where computational tasks are performed, increasing the CPU
speed.  Near the center of $C_0$, a {\tt fork-th} instruction is
executed initiating secondary execution (of the same code) at line 27,
giving rise to gene $C_2$.  The primary thread continues to line 55,
the $S$ gene, where genome size is calculated and the memory for its
offspring is allocated.  Next, the primary instruction pointer runs
into gene $R$, the copy loop, where replication occurs.  It is
executed once for each of the 99 instructions in the genome (hence its
dark color in the figure).  When this process is complete, it moves on
through gene $I_0$ shuffling numbers around, and re-enters gene $D$
for a final division.

During this time, the secondary thread executes gene $C_2$ computing a
few basic logical operations.  $C_2$ ends with a {\tt jump-f} (jump
forward) instruction that initially fails.  Passing through gene
$I_1$, numbers are shuffled within the thread and the jump at line 72
diverts the execution back to the beginning of the organism.  From
this point on, its execution loops through $C_1$ and $C_2$ for a total
of 10 times, using the results of each pass as inputs to the next,
computing different tasks each time.  Note that for this organism, the
secondary thread is never involved in replication.  Similar
overlapping patterns appear in natural organisms, particularly
viruses.  Fig.~\ref{mapfig}B exhibits a gene map of the phage
$\Phi$X174 containing portions of genetic code that are expressed
multiple times, each resulting in a distinct protein~\cite{MBOG}.
Studies of evolution in the overlapping genes of $\Phi$X174 and other
viruses have isolated the primary characteristic hampering evolution.
Multiple encodings in the same portion of a genome necessitate that
mutations be neutral (or beneficial) in their net effect over {\em
  all} expressions or they are selected against.  Fewer neutral
mutations result in a reduced variation and in turn slower adaptation.
It has been shown that in both viruses~\cite{MIYA78} and Avida
organisms~\cite{OFRIA99}, overlapping expressions have between 50 and
60\% of the variation of the non-overlapping areas in the same genome,
causing genotype space to be explored at a slower pace.

In higher organisms, multiple genes do develop that overlap in a
portion of their encoding, but are believed to be evolved out through
gene duplication and specialization, leading to improved
efficiency~\cite{KEESE92}.  Unfortunately, viruses and {\avida}
organisms are both subject to high mutation rates with no error
correction abilities.  This, in turn, causes a strong pressure to
compress the genome, thereby minimizing the target for mutations. As
this is an immediate advantage, it is typically seized, although it
leads to a decrease in the adaptive abilities of the population in the
long term.

\subsection{Environmental Influence on Differentiation}

\begin{figure}[tb]
 \centerline{\psfig{figure=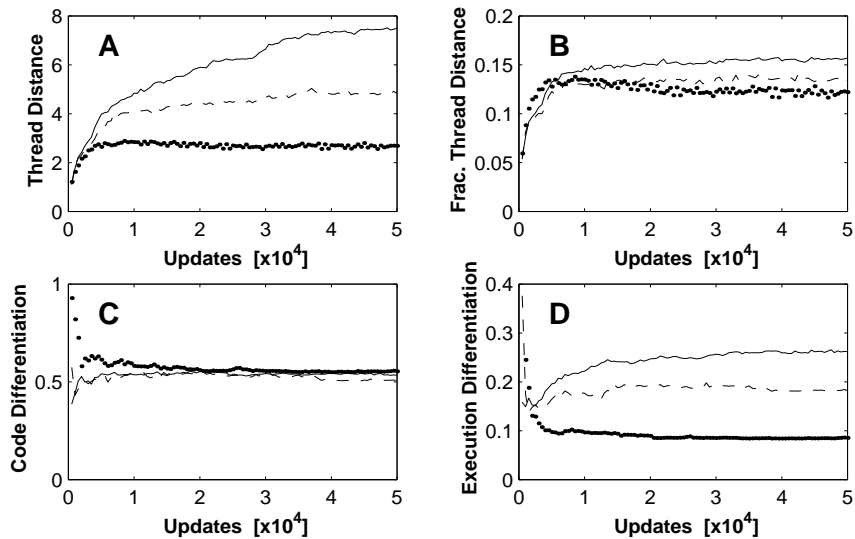,width=4.5in,angle=0}}
 \caption{ Differentiation measures averaged over all trials for each
   experiment. (A) Thread Distance, (B) Fractional Thread Distance,
   (C) Code Differentiation, (D) Expression Differentiation.
   Experiments from environment III (solid line), environment II
   (dashed line), and environment I (dotted line)}
 \label{difffig}
\end{figure}

Now that we have witnessed the development of multiple threads of
execution in {\avida}, let us examine the impact of environmental
complexity on this process.  Populations in all environments learn to
use their secondary thread quite rapidly, but show a marked difference
in their ability to diverge the threads into distinct functions.  In
Fig~\ref{difffig}A, average Thread Distance is displayed for all
trials in each environment showing a positive correlation between the
divergence of threads and the complexity of the environment they are
evolving in.

More complex environments provide more information to be stored within
the organism, promoting longer genomes~\cite{IAL98}, and possibly
biasing this measure.  To account for this, we consider this average
thread distance normalized to the length of the organisms, displayed
in Fig~\ref{difffig}B.  When threads fully differentiate, they often
execute neighboring sections of code, regardless of the length of the
genome they are in, biasing this measurement in the opposite
direction.  Longer genomes need their threads to be further spatially
differentiated in order to obtain an equivalent fractional thread
distance.  Thus, the fact that more complex environments give rise to
a marginally higher fractional distance is quite significant.

Interestingly, Code Differentiation (Fig~\ref{difffig}C) does not
firmly distinguish the environments, averaging at about 0.5.  In fact,
the distribution of code differentiation turns out to be nearly
uniform.  This indicates that the portion of the genomes that are
involved with the differentiated threads are similarly distributed
between complexity levels.  Execution Differentiation (the measure of
the fraction of executions that occurred differently between threads,
shown in Fig~\ref{difffig}D), however, once again positively
correlates environments with thread divergence.  The degree of
differentiation between the execution patterns is much more pronounced
in the more complex environments.

\section{Conclusions}

We have witnessed the development and differentiation of
multi-threading in digital organisms, and exhibited the role of
environmental complexity in promoting this differentiation.  Although
this is an inherently complex process, the ability to examine almost
any detail and dynamic within the framework of {\bf avida} provides
insight into what we believe are fundamental properties of biological
and computational systems.
 
The patterns of expression (lock-step, overlapping, and spatial
differentiation) are selected by balancing the ``physiological'' costs
of execution and differentiation against the implicit effects of
mutational load.  Clearly, multiple threads executing single regions
of the genome provides for additional use of that region.  The benefit
is in the form of additional functionality and a reduction in the
mutational load required for that functionality.  Within the context
of this thinking, the correlation between environmental complexity and
the usage of multiple threads makes a great deal of sense: multiple
threads are advantageous only if they can provide additional
functionality.

However, we have witnessed the cost side in this equation: when a gene
or gene product is used in multiple pathways, variations are reduced
as the changes to each gene must result in a net benefit to the
organism.  We observed a negative correlation between rates of
adaptation and use of multiple threads.  Furthermore, the ability to
analyze the entropy of each site in the genome quantifies the loss in
variability predicted by this hypothesis.  This entropy analysis has
been carried out in a biological context by Schneider~\cite{SSGE},
opening up opportunities to verify our results.

Implications of this work with potentially far reaching consequences
for Computer Science involve the study of {\it how} the individual
threads interact and what techniques the organisms implement to obtain
mutually robust operations.  The internal interactions within computer
systems lack the remarkable stability of biological systems to a
noisy, and often changing environment.  Life as we know it would never
have reached such vast multi-cellularity if every time a single
component failed or otherwise acted unexpectedly, the whole organism
shut down.
 
Clearly, we are still taking the first steps in developing systems of
computer programs that interact on similarly robust levels.  Here we
have performed experiments on a simple evolutionary system as a step
towards deciphering these biological principles as applied to digital
life.  In the future, we plan to add explicit costs for
multi-threading that depend on the {\em local} availability of
resources for thread execution.  Systems at levels of integration
anywhere near that of biological life are still a long way off, but
more concrete concepts such as applying principles from gene
regulation to develop self-scheduling parallel computers may be much
closer.

\vskip 0.5cm

\noindent
{\large {\bf Acknowledgements}}

\vskip 0.2cm
\noindent
This work was supported by the National Science Foundation under Grant
No. PHY-9723972.
G.H. was supported in part by a SURF fellowship from Caltech.
Access to a Beowulf system was provided by the Center for Advanced Computing
Research at the California Institute of Technology.  We would like to thank
an anonymous referee for useful comments.

\end{document}